# Eigenvalue analysis of a cantilever beam-rigid-body MEMS gyroscope

**Seyed Amir Mousavi Lajimi, Glenn Heppler, Eihab Abdel-Rahman**
Department of Systems Design Engineering, Faculty of Engineering,
University of Waterloo, 200 University Avenue West
Waterloo, Ontario, Canada, N2L 3G1
samousavilajimi@uwaterloo.ca, samousav@uwaterloo.ca

***Abstract*** **–**The eigenvalues of a new microbeam-rigid-body gyroscope are computed and studied to show the variation of frequencies versus the input spin rate. To this end, assuming the harmonic solution of the dynamic equation of motion the characteristic equation is obtained and solved for the natural frequencies of the system in the rotating frame. It is shown that the difference between the natural frequencies (eigenvalues) proportionally grows with the input angular displacement rate.

## 1. Introduction

Various designs and methods have been proposed to measure angular displacement, velocity, and acceleration by using MEMS gyroscopes. A conventional design of microgyroscopes includes a proof mass oscillating in two directions where the sense amplitude is employed to measure the rotation rate (Apostolyuk, 2006). To provide stable large vibration amplitude in the drive direction comb-drives with electrostatic actuation are used. Measuring the sense amplitude to estimate the input spin rate is the working principle in amplitude modulation vibratory microgyroscopes (Zotov et al., 2012). Zotov et al. (Zotov et al., 2012) introduced a method of estimating angular displacement rate based on measuring modal (natural) frequencies of the system.

We study the gyroscopic natural frequencies of the new cantilever beam-end-rigid-body (CBERB) microgyroscope to provide the essential information to operate the device in the frequency modulation mode. Performing initial studies on microbeams carrying end masses and end rigid bodies the authors (Lajimi et al., 2009; Lajimi and Heppler, 2012b; Lajimi and Heppler, 2012a; Lajimi and Heppler, 2013; Lajimi et al., 2013) have developed the governing equations of the microbeam gyroscope. To estimate the natural frequencies of the ideal system, the method of assumed modes is used to discretize the Lagrangian of the system. The discretized Lagrangian is substituted into Lagrange's equation and the reduce-order model is obtained and solved by assuming a harmonic solution.

## 2. Method

An eccentric end-rigid-body is attached to a uniform square microbeam that oscillates between two pairs of bottom and side electrodes, see Fig. 1. The total kinetic energy is given by

$$\begin{aligned} K &= K_B + K_M \\ &= \frac{1}{2}M\dot{\boldsymbol{R}}\cdot\dot{\boldsymbol{R}} + \dot{\boldsymbol{R}}\cdot\left(\boldsymbol{\omega}(L,t)\times\int_M \boldsymbol{\rho}\,\mathrm{d}M\right) \\ &\quad + \frac{1}{2}\boldsymbol{\omega}(L,t)\cdot\int_M \boldsymbol{\rho}\times(\boldsymbol{\omega}(L,t)\times\boldsymbol{\rho})\,\mathrm{d}M + \int_0^L m\,\dot{\boldsymbol{r}}_p\cdot\dot{\boldsymbol{r}}_p\,\mathrm{d}\xi \end{aligned} \tag{1}$$

where $M$ is the total mass of the end body, $\dot{\boldsymbol{R}}$ is the velocity vector of a reference point in the end-rigid-body chosen to coincide with end of the beam and computed relative to the base frame, $\boldsymbol{\omega}(L,t)$ is the angular velocity of the beam's section at $L$, $\boldsymbol{\rho}$ is the position vector of an arbitrary point in the end rigid body relative to the end of the beam, $m$ is the mass per unit length of the beam, $p$ is an arbitrary point on the beam's cross-section, and $\dot{\boldsymbol{r}}_p$ is the velocity vector for the point $p$ relative to the inertial base frame, see Fig.1.

Computing the elastic and electrostatic potential energy and substituting the energy expressions into the Lagrangian, $L = K - P$, nondimensionalizing the Lagrangian, and assuming the response of the beam $w(\hat{\xi},\hat{t})$ and $v(\hat{\xi},\hat{t})$ in generalized coordinates

$$w(\hat{\xi},\hat{t}) = w_s(\hat{\xi}) + \psi(\hat{\xi})\, q(\hat{t}) \quad \text{and} \quad v(\hat{\xi},t) = v_s(\hat{\xi}) + \phi(\hat{\xi})\, p(\hat{t}) \tag{2}$$

a single-mode approximation of the Lagrangian is obtained. The Lagrangian is substituted in Lagrange's equation and reduced-order model is obtained. The equation of motion is linearized about the static equilibrium configuration as

$$\begin{aligned}
&\left(\Lambda + \alpha\Lambda' + \widehat{M}\phi(1)^2 + 2\,\hat{e}\,\widehat{M}\,\phi(1)\,\phi'(1) + \hat{J}_{33}(1)^2 + \widehat{M}\hat{e}^2\phi'(1)^2\right)\ddot{p}(\hat{t}) + c\Lambda\dot{p}(\hat{t})\\
&\quad + p(\hat{t})\left(\Lambda'' + \Omega^2(\alpha\,\Lambda' - \Lambda) - \widehat{M}\,\Omega^2\big(\phi(1) + \hat{e}\phi'(1)\big)^2\right.\\
&\quad \left. - \Omega^2\big(\hat{J}_{22} - \hat{J}_{11}\big)\phi'(1)^2\right)\\
&\quad - \Omega\,\dot{q}(\hat{t})\,\big(2\,\Pi\\
&\quad + 2\,\widehat{M}\,\big(\hat{e}\,\psi(1)\phi'(1) + \hat{e}\,\phi(1)\psi'(1) + \phi(1)\psi(1) + \hat{e}^2\,\phi'(1)\psi'(1)\big)\\
&\quad + \big(\hat{J}_{33} + \hat{J}_{22} - \hat{J}_{11}\big)\,\phi'(1)\psi'(1)\,\big)\\
&\quad = \frac{2\,\hat{e}\,\nu\,V_{\mathrm{DC}}^2}{(1 - v_s(1) - \hat{e}\;\;{v_s}'(1))^3}\big(\phi(1) + \hat{e}\,\phi'(1)\big)^2 p(\hat{t})
\end{aligned} \tag{3}$$

where

$$\begin{gathered}
\Lambda = \int_0^1 \phi(\hat{\xi})^2\,\mathrm{d}\hat{\xi}, \qquad \Lambda' = \int_0^1 \phi'(\hat{\xi})^2\mathrm{d}\hat{\xi}, \qquad \Gamma = \int_0^1 \psi(\hat{\xi})^2\,\mathrm{d}\hat{\xi},\\
\Gamma' = \int_0^1 \psi'(\hat{\xi})^2\mathrm{d}\hat{\xi}, \quad \Pi = \int_0^1 \phi(\hat{\xi})\,\psi(\hat{\xi})\,\mathrm{d}\hat{\xi}
\end{gathered} \tag{4}$$

and the nondimensional parameters are defined as

$$\begin{gathered}
\hat{\xi} = \frac{\xi}{L}, \qquad \hat{t} = \frac{t}{\kappa}, \qquad \hat{e} = \frac{e}{L}, \qquad \nu = \frac{6\,\epsilon\,h\,L^4}{E\,b^4 g^3}, \qquad \widehat{M} = \frac{M}{\rho\,a\,b\,L}, \;\; \hat{J}_{11} = \frac{J_{11}}{\rho\,a\,b\,L^3},\\
\hat{J}_{22} = \frac{J_{22}}{\rho\,a\,b\,L^3}, \qquad \hat{J}_{11} = \frac{J_{22}}{\rho\,a\,b\,L^3}, \qquad \kappa = \frac{12\rho L^4}{E\,b^2}, \qquad \alpha = \frac{b^2}{12\,L^2}, \quad \widehat{\Omega} = \kappa\,\Omega.
\end{gathered} \tag{5}$$

Mode shapes $\psi(x)$ and $\phi(x)$ are obtained by setting the input angular rate to zero, $\Omega = 0$, in the equation of motion and solving the frequency equation for the uncoupled system. A similar equation is obtained for the drive direction, however $V_{\mathrm{DC}}$ is replaced with $V_{\mathrm{DC}} + V_{\mathrm{AC}}$. To compute the natural frequencies it is assumed that

$$q(\hat{t}) = z\,e^{s\,\hat{t}} \quad \text{and} \quad p(\hat{t}) = y\,e^{s\,\hat{t}} \tag{6}$$

Substituting Eq. (5) into (3) and its counterpart in the drive direction, collecting eigenvectors $z$ and $y$, and setting the determinant of the coefficient matrix equal to zero, the characteristic equation is obtained as

$$C_4 s^4 + C_2 s^2 + E_4 \Omega^4 + E_2 \Omega^2 + E_0 = 0 \qquad (7)$$

where $C_4, C_2, E_4, E_2$, and $E_0$ are functions of the system parameters and mode shapes, $\psi(x)$ and $\phi(x)$, of the uncoupled governing equations of motion ($\Omega = 0$). By solving Eq. (6) for $s$ the undamped eigenvalues or natural frequencies of the system are calculated. Equation (7) is used to identify damped frequencies and the instability region due to the angular rotation rate. To this end, imaginary and real parts of Eq. (7) are set to zero independently.

The effects of DC loading can also be included in Eq.(7) by dropping terms associated with AC loading in Eq. (3), i.e. setting $V_{AC} = 0$, but keeping terms associated with DC loading, i.e. $V_{AC} \neq 0$, and following a very similar procedure to the previous approach. Furthermore, by algebraically manipulating the solutions of Eq. (6) a closed-form relation for computing the angular spin rate Ω in terms of the modal or natural frequencies is obtained. The aforementioned result is significant in that the input angular rotation rate is estimated based on modal frequency measurement. Zotov et al. (Zotov et al., 2011; Zotov et al., 2012) have reported similar results for mass-spring microgyroscopes.

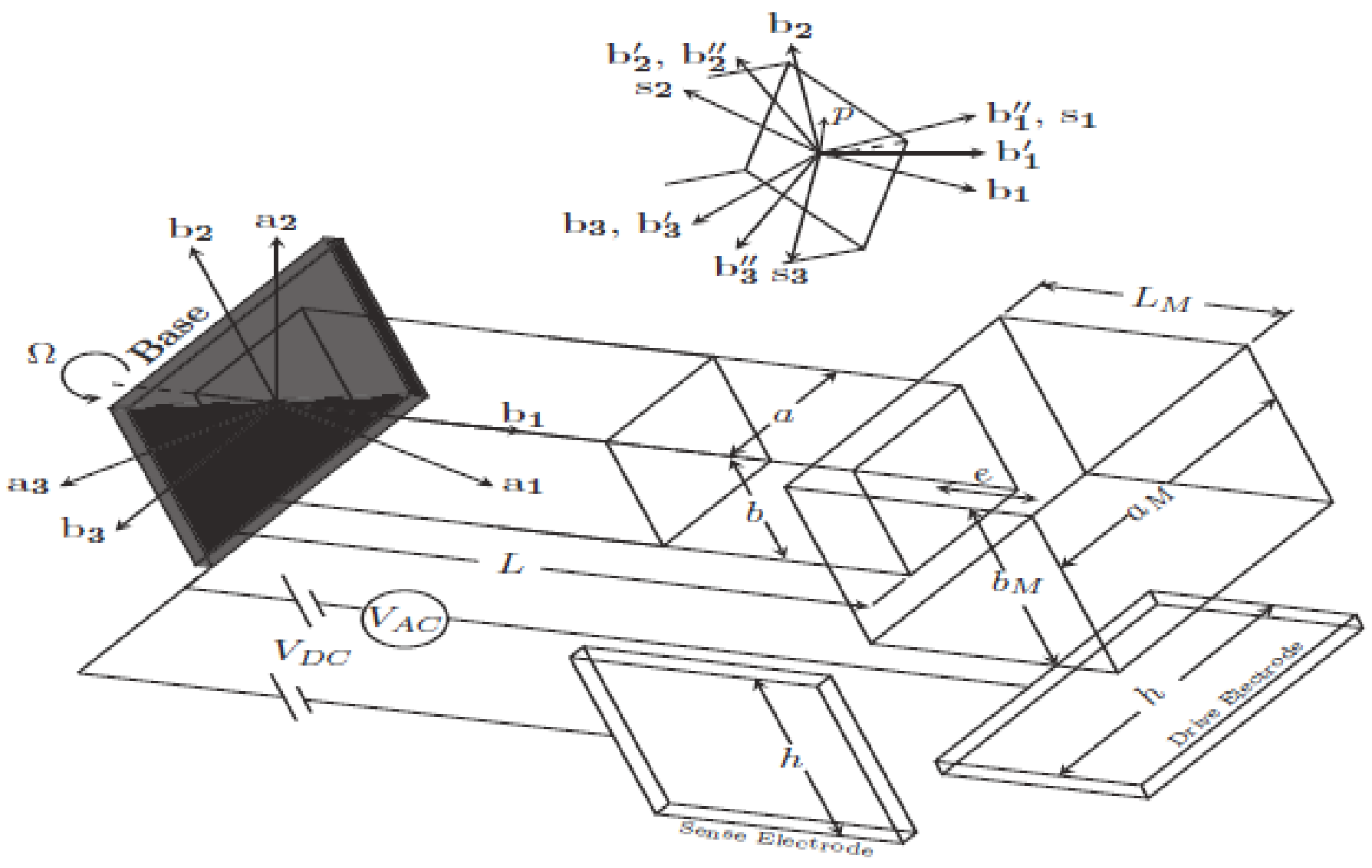

Fig. 1 The microgyroscope: the gyroscope rotates about longitudinal axis. Systems of inertial $(a_1, a_2, a_3)$, base $(b_1, b_2, b_3)$, and sectional $(s_1, s_2, s_3)$ frames are used to obtain the mathematical model of MEMS gyroscope. The eccentricity (the distance between beam's end and end-rigid-body's center of mass) is denoted by $e$.

Table 1 System parameters

| $L$ ($\mu$m) | $a$ ($\mu$m) | $b$ ($\mu$m) | $L_M$ ($\mu$m) | $a_M$ ($\mu$m) | $b_M$ ($\mu$m) | $h$ ($\mu$m) | $g$ ($\mu$m) | $e$ ($\mu$m) | $E$ (GPa) | $\rho$ $(\frac{\text{kg}}{\text{m}^3})$ | $\epsilon$ $(\frac{F}{m})$ |
|---|---|---|---|---|---|---|---|---|---|---|---|
| 400 | 10 | 10 | 100 | 200 | 10 | 10 | 2 | 50 | 160 | 2330 | $8.854 \times 10^{-12}$ |

## 3. Numerical Results

By setting $\Omega = 0$ in Eq. (3) the uncoupled set of dynamic equations are obtained and the characteristic equation for each direction is computed separately. Solving the characteristic equations the first two natural frequencies are computed and plotted in Figs. 2 and 3. The dashed lines and solid lines represent the natural frequency in the sense and drive directions respectively. The fundamental frequency, Fig. 2, decreases to zero as the DC loading approaches the static pull-in voltage, $V_{\text{PI}} = 35.263\text{V}$. Figure 3 shows the variation in the second natural frequency of uncoupled system ($\Omega = 0$). It is expected to see an insignificant reduction in the second natural frequency just before pull-in. Mode shapes are computed for the uncoupled motion of the structure in the sense and drive directions. In Fig. 4 the gyroscopic natural frequencies (roots of characteristic equation, Eq. (7)) are plotted vs. the input rotation rate for $V_{\text{DC}} = 0\text{V}$. It is noted that at zero input rate, $\Omega = 0$, the natural frequencies are not equal to each other and diverge from each other as the input angular rotation rate increases a typical characteristic of a gyroscopic system.

Altering the DC loading the natural frequency branches either approach each other or diverge from each other. Moreover, the slope of frequency curves varies with the varying bias voltage. In Fig. 5 the bifurcation diagram of natural frequencies is presented for $V_{\text{DC}} = 10$ V in the drive direction and $V_{\text{DC}} = 5$ V in the sense direction. It is seen that the frequency curves have approached each other which corresponds to moving along the frequency curve in Fig. 2. Various methods have been proposed to measure input angular rotation rate including the frequency modulation method. To measure the input rotation rate based on the frequency modulation method, the natural frequency split is computed (Zotov et al., 2012). To this end, a closed-form relation relating input angular rate to the system parameters and modal frequencies is obtained. In Figs. 6 and 7 the variation of differential modal frequency measurement with input spin rate are plotted for the same two DC voltage cases used in Figs. 4 and 5. The higher slope in Fig. 7 suggests that using the sensor in a mode-matched configuration results in higher sensitivity of the microgyrosocpe. In practice, modal frequencies are measured and the input angular rate is computed based on the difference between the modal frequencies.

## 4. Conclusion

We have studied the bifurcation of natural frequencies due to the input angular rotation rate. The angular displacement rate appeared in the form of Coriolis and Centrifugal forces in the equation of motion and caused the nonlinear evolution of the natural (modal) frequencies. It is shown that the frequency modulation method can be used to measure angular velocity based on differential natural frequency measurement.

## References

Apostolyuk, V. (2006). Theory and design of micromechanical vibratory gyroscopes. Mems/nems, Springer**:** 173-195.

Lajimi, A. M., Abdel-Rahman, E. and Heppler, G. R. (2009). On natural frequencies and mode shapes of microbeams. Hong Kong, Int Assoc Engineers-IAENG.

Lajimi, S. A. M. and Heppler, G. R. (2012a). "Comments on “natural frequencies of a uniform cantilever with a tip mass slender in the axial direction”." Journal of Sound and Vibration **331**(12): 2964-2968.

Lajimi, S. A. M. and Heppler, G. R. (2012b). Eigenvalues of an axially loaded cantilever beam with an eccentric end rigid body. Proceedings of International Conference on Mechanical Engineering and Mechatronics, August 16-18, Ottawa, Canada**:** 135-131-135-138.

Lajimi, S. A. M. and Heppler, G. R. (2013). "Free vibration and buckling of cantilever beams under linearly varying axial load carrying an eccentric end rigid body." Transactions of the Canadian Society for Mechanical Engineering **Accepted for publication**.

Lajimi, S. A. M., Heppler, G. R. and Abdel-Rahman, E. (2013). A new cantilever beam rigid-body mems gyroscope: Mathematical model and linear dynamics. Proceedings of 2nd International Conference on Mechanical Engineering and Mechatronics, August 8-9, Toronto, Canada**:** 177-171-177-176.

Zotov, S. A., Prikhodko, I. P., Trusov, A. A. and Shkel, A. M. (2011). Frequency modulation based angular rate sensor. Proceedings of Micro Electro Mechanical Systems (MEMS), 2011 IEEE 24th International Conference on**:** 577-580.

Zotov, S. A., Trusov, A. A. and Shkel, A. M. (2012). "High-range angular rate sensor based on mechanical frequency modulation." Journal of Microelectromechanical Systems **21**(2): 398-405.

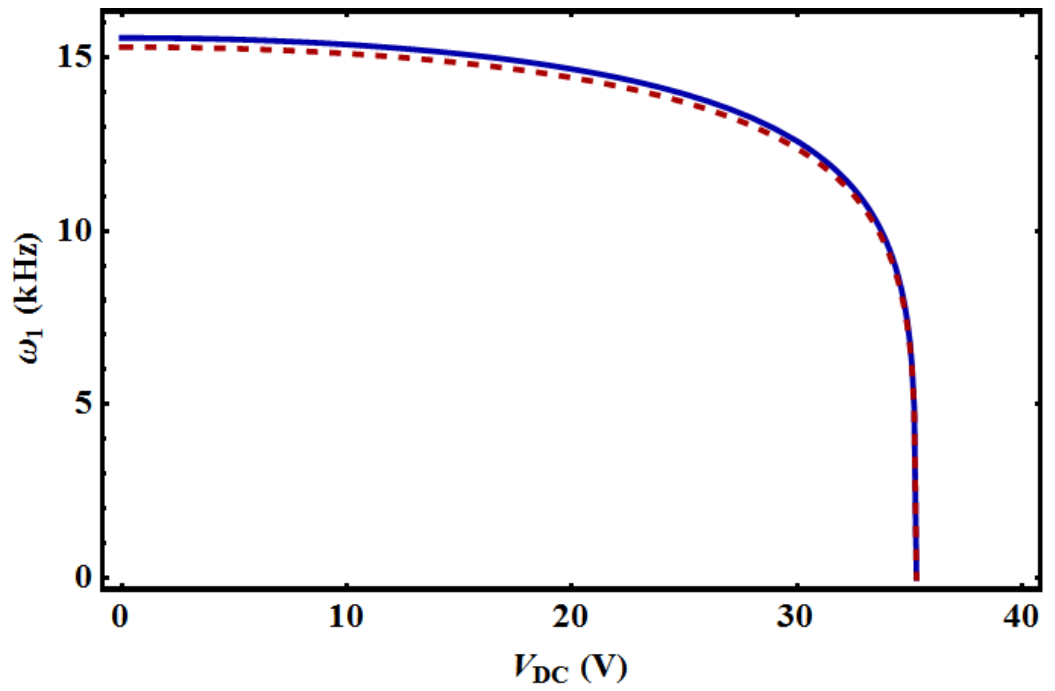

Fig. 2 The variation of the first natural frequency vs. DC loading for the uncoupled system

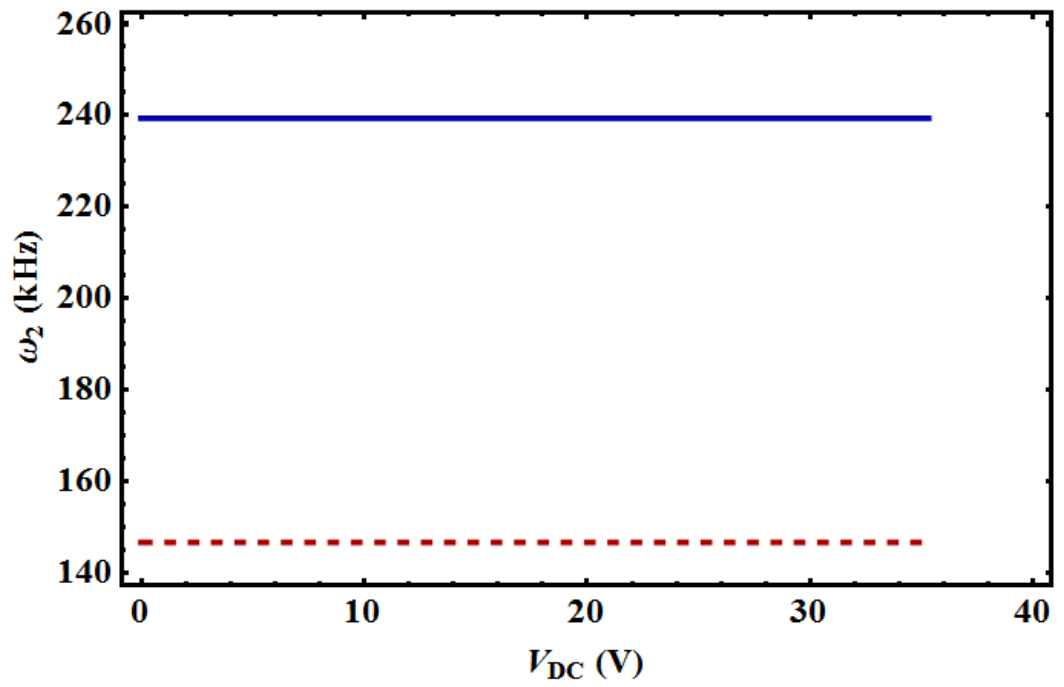

Fig. 3 The variation of the second natural frequency vs. DC loading for the uncoupled system

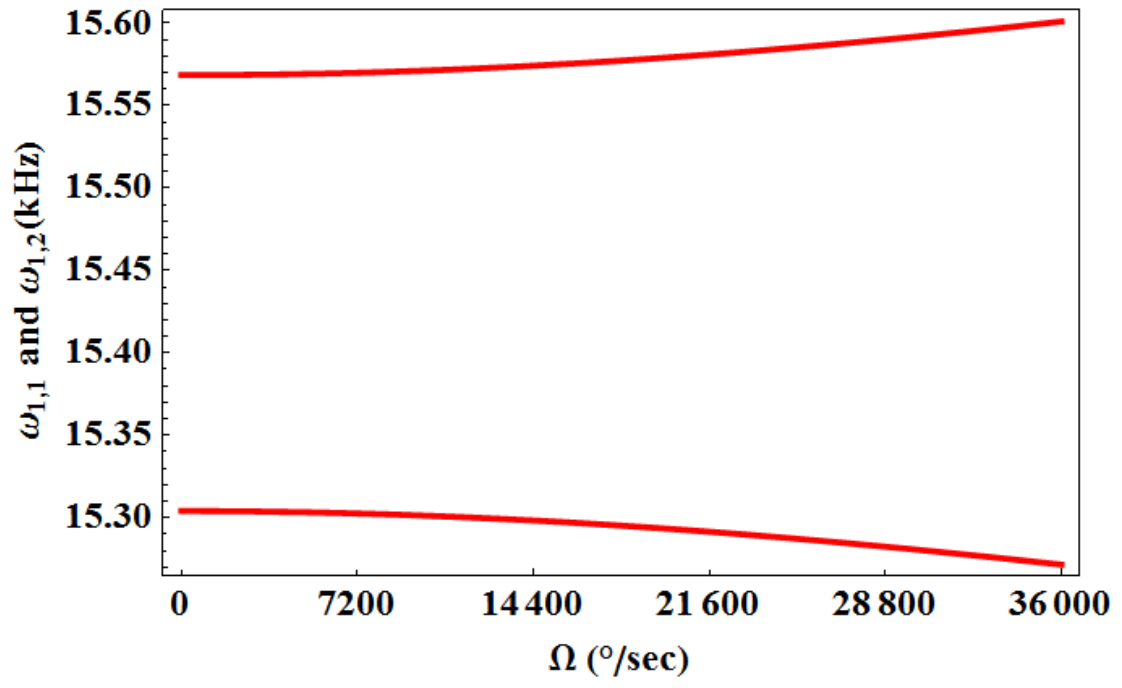

Fig. 4 Gyroscopic natural frequencies; $V_{\mathrm{DC}} = 0\mathrm{V}$

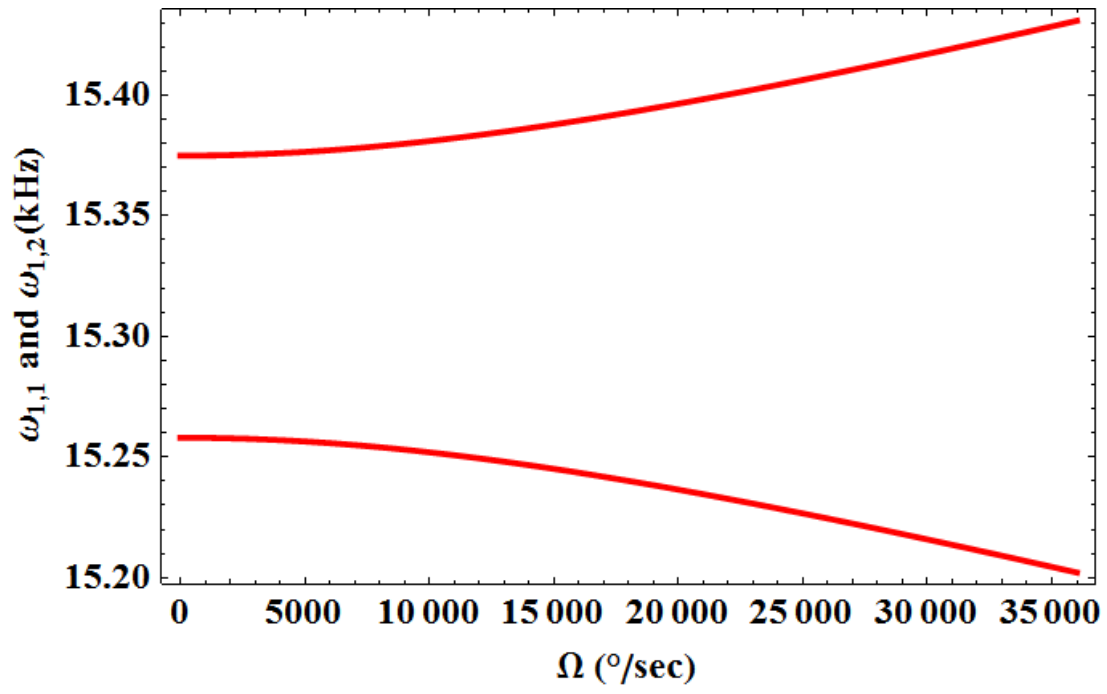

Fig. 5 Gyroscopic natural frequencies; $V_{\mathrm{DC}}$ =10 V in the drive and $V_{\mathrm{DC}}$=5 V in the sense directions

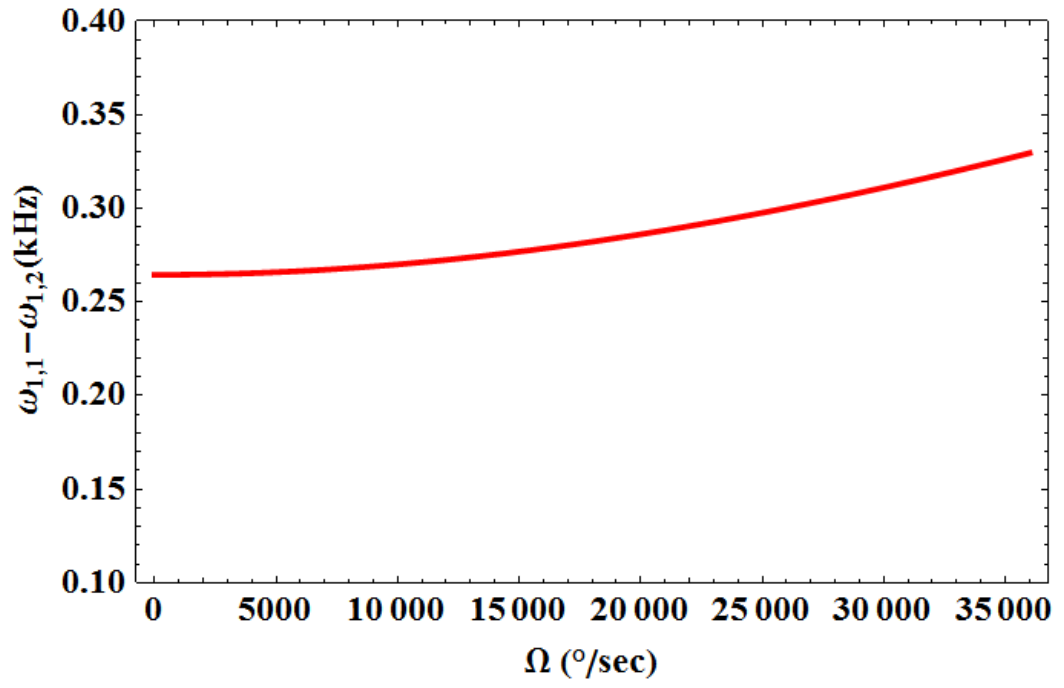

Fig. 6 The frequency split for the first natural frequency; $V_{\mathrm{DC}} = 0\mathrm{V}$ in both sense and drive directions

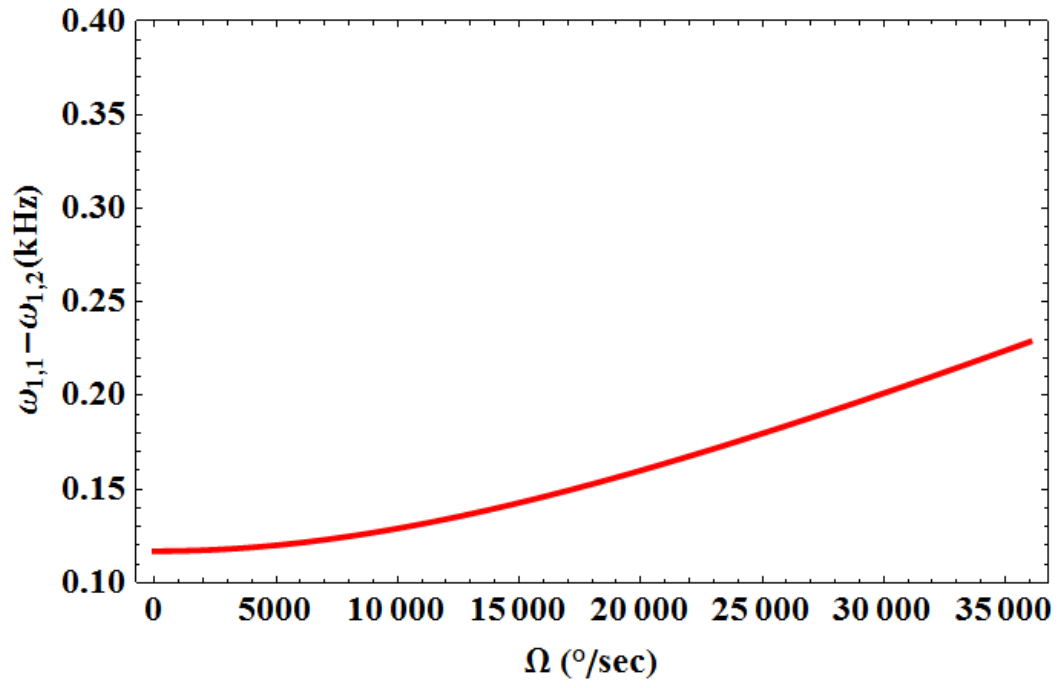

Fig.7 The frequency split for the first natural frequency; $V_{\mathrm{DC}}$=10 V in the drive and $V_{\mathrm{DC}}$=5 V in the sense directions